\documentclass[prb,aps,twocolumn,twoside,notitlepage]{revtex4-1}

\usepackage{amsmath,amsfonts,amssymb,amsthm,mathtools}
\usepackage{xcolor}
\usepackage[english]{babel}
\usepackage{latexsym}
\usepackage[pdftex]{graphicx}
\usepackage{epstopdf}
\usepackage{subfigure}
\newcommand{\vu}{\textbf{u}}

\newcommand{\vvu}{\langle\langle\textbf{u}\rangle\rangle}
\newcommand{\LL}{\left\langle}
\newcommand{\RR}{\right\rangle}

\newcommand{\x}{\textbf{x}}
\newcommand{\w}{\boldsymbol{\omega}}

\newcommand{\curl}{\boldsymbol{\nabla}\times}
\newcommand{\grad}{\boldsymbol{\nabla}}

\newcommand*{\citen}[1]{%
  \begingroup
    \romannumeral-`\x 
    \setcitestyle{numbers}%
    \cite{#1}%
  \endgroup   
}

\begin{document}
\title{Emergence of vorticity and viscous stress in finite-scale quantum hydrodynamics}
\author{Christopher Triola}
\affiliation{Los Alamos National Laboratory, Los Alamos, New Mexico 87545, USA}
\date{\today}
\begin{abstract}
The Madelung equations offer a hydrodynamic description of quantum systems, from single particles to quantum fluids. In this formulation, the probability density is mapped onto the fluid density and the phase is treated as a scalar potential generating the velocity field. As examples of potential flows, quantum fluids described in this way are inherently irrotational, but quantum vortices may arise at discrete points where the phase is undefined. In this paper, starting from this irrotational description of a quantum fluid, a coarse-graining procedure is applied to arrive at a macroscopic description of the quantum fluid in terms of a hierarchy of moments in which the role of velocity is played by a Favre average of the microscopic velocity field. This hierarchy is truncated using an explicit closure derived from an expansion in a finite length scale. The resulting coarse-grained fields are shown to allow for finite vorticity at any point in the fluid. Furthermore, it is shown that this vorticity obeys a similar equation to the vorticity equation in classical hydrodynamics and includes a vortex-stretching term. The particular closure employed here also gives rise to a novel stress term in the fluid equations, which in the appropriate limit appears analogous to an artificial viscous stress from computational fluid dynamics.

\end{abstract}
\maketitle

Fluid turbulence is known to be a ubiquitous non-equilibrium phenomenon arising in fluids when the kinetic energy injected at large scales dominates the dissipation of energy at small scales due to viscosity\cite{mccomb1990physics,mccomb1995theory,sreenivasan1999fluid,eyink2006onsager}. While first identified and extensively studied in classical fluids, turbulence has also been observed in quantum fluids, including both superfluids and ultra-cold atomic gases\cite{vinen2002quantum,barenghi2009vortex,vinen2010quantum,paoletti2011quantum,tsubota2014turbulence,barenghi2014introduction}. While there is no consensus on the precise definition of turbulence, it is characterized by spatially complex and temporally aperiodic flow fields and involves processes on many length scales. In three-dimensions, it is well established that turbulence involves a cascade of energy from large to small length scales\cite{kolmogorov1991local,falkovich2009symmetries,dubrulle2019beyond,carbone2020vortex}. While the details of this process are not fully understood, there is significant evidence that, in classical turbulence, this cascade is related to the non-linear phenomena of strain self-amplification and vortex stretching\cite{carbone2020vortex}. In quantum turbulence, this cascade has a remarkably similar behavior at large length scales, with the Kolmogorov spectrum being observed in superfluid $^4$He and simulations using the Non-linear Schr\"{o}dinger equation (NLSE)\cite{paoletti2011quantum}. 
However, in contrast to the case of classical turbulence where the cascade is related to the dynamics of a continuous vorticity field, the cascade in quantum turbulence is related to the dynamics of a discrete number of quantized vortices. The differences between these two cases are most pronounced at length scales less than the inter-vortex spacing, where significant contributions to the energy cascade have their origin in transverse modes (Kelvin waves) which propagate along the quantized vortices\cite{barenghi2009vortex}. The two cases appear much more similar at larger length scales where the dynamics are dominated by the collective motion of many vortices.  
While this similarity between classical and quantum turbulence is not unexpected based on intuition from the correspondence principle\cite{paoletti2011quantum}, a full theory establishing the rigorous connection between these two regimes is lacking, due in part to the lack of a single governing equation used to describe quantum turbulence at all length scales. 

In contrast to the study of classical turbulence, whose phenomenology is believed to emerge from the dynamics governed by a single equation, the Navier-Stokes equation, current descriptions of quantum turbulence require a hierarchy of models\cite{barenghi2014introduction}. The microscopic behavior of the quantum fluid is modeled using the NLSE, or appropriate generalizations to capture strong interactions; the mesoscale behavior is described by vortex-filament models; and macroscopic dynamics are described using the Hall–Vinen–Bekharevich–Khalatnikov (HVBK) model\cite{barenghi2014introduction, zhang2023higher}. Previous works have investigated the coarse-grained dynamics of many-vortex solutions of the NLSE, either by deriving an emergent non-Eulerian vortex-fluid hydrodynamics from the point-vortex model\cite{yu2017emergent}, or by developing kinetic and statistical theories based on coarse-grained vortex distributions\cite{salman2016long}. The quantum fluid equations have also been coarse-grained directly, in a work by Tanogami, using a spatial filter to derive a hierarchy of filtered moment equations\cite{tanogami2021theoretical}. A key purpose of the present paper is to extend the coarse-graining framework introduced by Tanogami by providing a closure of the hierarchy of filtered moment equations. This closure yields explicit expressions for the emergent sub-scale stress terms that appear in the unclosed formulation. In addition, we demonstrate, using a family of simple analytic examples, that the coarse-grained velocity field in three dimensions can exhibit finite, continuous vorticity, and that this coarse-grained vorticity obeys an evolution equation closely analogous to the classical vorticity equation. The emergence of a novel viscous stress term due to integrating out small scale degrees of freedom has similarities with a recent work which demonstrated a relationship between decoherence and viscosity\cite{zhou2025emergent}.

The particular closure introduced in this paper is based on finite scale theory, a method previously studied in the context of classical continuum dynamics\cite{margolin2002rationale,margolin2009finite,margolin2014finite,margolin2019reality}.
The motivation for finite scale theory in classical continua is based on the fact that in order to solve many problems in classical continuum dynamics, the continuum equations are almost always approximated by discretizing the domain and forming a mesh, represented by a finite number of points in the original space. In the context of Lagrangian hydrocodes, this discretization amounts to an averaging of the continuum variables over finite size cells. However, these averaged quantities are not necessarily solutions to the original partial differential equations governing the dynamics of the continuum. Finite scale theory provides a connection between the coarse-graining of the continuous domain and the differential equations describing the averaged quantities.  

The remainder of this paper is organized as follows. In Sec.~\ref{sec:madelung} the quantum fluid equations are introduced explicitly, highlighting their similarities to classical fluid equations. Sec.~\ref{sec:CG} contains a pedagogical derivation of the coarse-graining procedure which will be applied to the quantum fluid equations as well as a derivation of the closure from finite scale theory which will be used to truncate the hierarchy of filtered moments. In Sec.~\ref{sec:FS} the coarse-graining procedure and closure model are applied to the Madelung equations and both the general (unclosed) and finite-scale (closed) equations are presented and discussed. In Sec.~\ref{sec:vorticity} the equation governing the evolution of vorticity is derived based on the results in the previous sections. Sec.~\ref{sec:example} presents a family of simple examples in which an irrotational velocity field gives rise to a finite and continuous vorticity field after undergoing the coarse-graining procedure discussed in the preceding sections. Sec.~\ref{sec:conclusions} provides a discussion of the conclusions that can be drawn from both the general equations as well as the simple examples.

\section{Madelung Equations}
\label{sec:madelung}
Consider a complex wavefunction $\psi$, whose dynamics are governed by the non-linear Schr\"{o}dinger equation. Rewriting the complex function in terms of its magnitude and phase, $\psi=\sqrt{\tfrac{1}{m}\rho} \ e^{\tfrac{i}{\hbar}\theta}$, the equations governing the evolution of $\rho$ and $\theta$ may be written as:
\begin{equation}
\begin{aligned}
\partial_t \rho + \grad \cdot \left( \rho \textbf{u} \right) &=0, \\
\partial_t \vu + \left( \vu\cdot\grad \right) \vu &= \frac{\hbar^2}{2m^2} \grad \left( \frac{\nabla^2 \sqrt{\rho}}{\sqrt{\rho}}\right)- \frac{g}{m}\grad\rho,
\end{aligned}
\label{eq:madelung}
\end{equation}
where we define $\vu\equiv\grad\theta/m$. These are the Madelung equations\cite{madelung1927quantum} which offer a hydrodynamic interpretation to the wavefunction due to their similarity to the continuity equation and Cauchy momentum equation of fluid mechanics:
\begin{equation}
\begin{aligned}
\partial_t \rho + \grad \cdot \left( \rho \textbf{u} \right) &=0, \\
\partial_t \vu + \left( \vu\cdot\grad \right) \vu &= \frac{1}{\rho} \grad\cdot \boldsymbol{\sigma} + \textbf{f},
\end{aligned}
\label{eq:hydro}
\end{equation}
where $\rho$ is the mass density of the continuum, $\vu$ is the velocity field, $\boldsymbol{\sigma}$ is the stress tensor, and $\textbf{f}$ is an external force. When $\grad \cdot \boldsymbol{\sigma} = -\grad p$, the above momentum equation becomes the Euler equation, when $\grad \cdot \boldsymbol{\sigma} = -\grad p + \mu \nabla^2 \vu +\tfrac{1}{3}\mu \grad \left( \grad\cdot \vu\right)$, it becomes the Navier-Stokes equation.

Ignoring the physical interpretations of these equations, it is clear that Eqs. (\ref{eq:madelung}) are simply a special case of Eqs. (\ref{eq:hydro}), with the choice:
$$\frac{1}{\rho} \grad\cdot \boldsymbol{\sigma} + \textbf{f} =  \frac{\hbar^2}{2m^2} \grad \left( \frac{\nabla^2 \sqrt{\rho}}{\sqrt{\rho}}\right)- \frac{g}{m}\grad\rho. $$
However, it is worth emphasizing that, by definition, the velocity appearing in the Madelung equations is given by the gradient of the phase, $\vu\equiv\grad\theta/m$ and therefore the vorticity, $\w=\curl\vu$, must vanish except at singular points in the phase. This is not the case in the classical fluid dynamics described by Eqs. (\ref{eq:hydro}). In fact, as mentioned above, the vorticity is believed to play an important role in the Kolmogorov cascade in classical turbulence.

\section{Generalized Finite Scale Analysis}
\label{sec:CG}
Consider a continuum field $A(\textbf{x},t)$ where $\textbf{x}$ and $t$ are space and time coordinates, respectively. Let us define the coarse-grained field as:
\begin{equation}
\langle A(\textbf{x},t) \rangle \equiv \int_{-\infty}^{\infty}\int_{-\infty}^{\infty}\int_{-\infty}^{\infty} d^3x' f(\textbf{x}') A(\textbf{x}+\textbf{x}',t), 
\label{eq:ave}
\end{equation}
where $f$ is a normalized distribution describing the details of our coarse-graining procedure. At this level, the spatial filtering procedure here is identical to the one employed by Tanogami\cite{tanogami2021theoretical} in his analysis of the energy cascade in quantum fluids. 

The procedure employed in previous studies of finite scale closures would correspond to the choice of a uniform box distribution\cite{margolin2002rationale,margolin2009finite}; however, in principle, many other choices could yield similar results. For simplicity, we will assume that $f$ is a separable distribution, $f(\textbf{x})=f_1(x_1)f_2(x_2)f_3(x_3)$ and that each function $f_i(x)$ is an even function of the variable $x$. In this section we will derive the general expressions without making further assumptions about the functional form of $f$ and defer a discussion of reasonable choices for $f$ to a later section.

Throughout this work, we denote the moments of this distribution as:
\begin{equation}
\mu_{n_1, n_2, n_3} \equiv \int_{-\infty}^{\infty} \int_{-\infty}^{\infty} \int_{-\infty}^{\infty} d^3 x f(\textbf{x}) x_1^{n_1}x_2^{n_2}x_3^{n_3} .  
\label{eq:moments}
\end{equation}
Note that, because $f$ is a separable distribution, these moments are separable: $\mu_{n_1,n_2,n_3}=\mu_{1,n_1}\mu_{2,n_2}\mu_{3,n_3}$, where $\mu_{i, n} \equiv \int_{-\infty}^{\infty} d x f_i(x) x^{n}$. Moreover, because we have assumed that $f$ is even in each of the spatial coordinates, all odd moments are zero, $\mu_{i,2n+1}=0$. 

With these conventions, and assuming that $A(\textbf{x},t)$ is sufficiently smooth in the neighborhood of $\textbf{x}$ that it may be expanded in a Taylor series, it is straightforward to derive the following relations:
\begin{widetext}
\begin{align*}
\langle A (\textbf{x},t)\rangle &= \sum_{n_1,n_2,n_3=0}^{\infty} \frac{\mu_{n_1,n_2,n_3}}{n_1! n_2! n_3!} \partial_{x_1}^{n_1}\partial_{x_2}^{n_2}\partial_{x_3}^{n_3} A(\textbf{x},t) , \\
\langle \partial_{x_1}^{m_1} \partial_{x_2}^{m_2}\partial_{x_3}^{m_3} A (\textbf{x},t)\rangle &= \sum_{n_1,n_2,n_3=0}^{\infty} \frac{\mu_{n_1,n_2,n_3}}{n_1! n_2! n_3!} \partial_{x_1}^{n_1+m_1}\partial_{x_2}^{n_2+m_2}\partial_{x_3}^{n_3+m_3} A(\textbf{x},t)=\partial_{x_1}^{m_1} \partial_{x_2}^{m_2}\partial_{x_3}^{m_3}\langle  A (\textbf{x},t)\rangle. \\
\end{align*}
Note that the averaging commutes with spatial derivatives. Now, transform to dimensionless variables, $\xi_i=x_i/L_i$:
\begin{align*}
\frac{1}{L_{1}^{m_1} L_{2}^{m_2}L_{3}^{m_3}} \langle  \partial^{m_1}_{\xi_1}\partial^{m_2}_{\xi_2}\partial^{m_3}_{\xi_3} A (\boldsymbol{\xi},t)\rangle &= \sum_{n_1,n_2,n_3=0}^{\infty} \frac{1}{n_1! n_2! n_3!} \frac{\mu_{n_1,n_2,n_3}}{L_{1}^{n_1+m_1} L_{2}^{n_2+m_2}L_{3}^{n_3+m_3}} \partial_{\xi_1}^{n_1+m_1}\partial_{\xi_2}^{n_2+m_2}\partial_{\xi_3}^{n_3+m_3} A(\boldsymbol{\xi},t), \\
\rightarrow \langle  \partial^{m_1}_{\xi_1}\partial^{m_2}_{\xi_2}\partial^{m_3}_{\xi_3} A (\boldsymbol{\xi},t)\rangle &= \sum_{n_1,n_2,n_3=0}^{\infty} \frac{1}{n_1! n_2! n_3!} \frac{\mu_{n_1,n_2,n_3}}{L_{1}^{n_1} L_{2}^{n_2}L_{3}^{n_3}} \partial_{\xi_1}^{n_1+m_1}\partial_{\xi_2}^{n_2+m_2}\partial_{\xi_3}^{n_3+m_3} A(\boldsymbol{\xi},t), \\
\end{align*}
\end{widetext}
where $L_i$ is a relevant macroscopic length scale. Note that $\mu_{n_1,n_2,n_3}\sim \ell_1^{n_1}\ell_2^{n_2}\ell_3^{n_3}$, where $\ell_i$ are characteristic length scales associated with the spatial dependence of the distribution $f$. We now assume that each of these scales, $\ell_i$, is small compared to the corresponding macroscopic scale $L_i$:
\begin{equation}
\frac{\ell_i}{L_i} \equiv\eta_i<<1. 
\end{equation} 
When this condition is satisfied, we may truncate the infinite series above:
\begin{align*}
    \langle A (\boldsymbol{\xi},t)\rangle &= A(\boldsymbol{\xi},t) + \frac{1}{2}\sum_{i=1}^3\frac{\mu_{i,2}}{L_{i}^{2}} \partial_{\xi_i}^{2} A(\boldsymbol{\xi},t) +\mathcal{O}(\eta_i^4), \\
    \langle \partial_{\xi_k}^{2} A (\boldsymbol{\xi},t)\rangle &= \partial_{\xi_k}^{2} A (\boldsymbol{\xi},t) + \frac{1}{2}\sum_{i=1}^3\frac{\mu_{i,2}}{L_{i}^{2}} \partial_{\xi_i}^{2} \partial_{\xi_k}^{2} A (\boldsymbol{\xi},t)  +\mathcal{O}(\eta_i^4). \\
\end{align*}
Collecting the terms above and converting back to dimensionful coordinates, we can rewrite the function $A$ in terms of coarse-grained quantities:
\begin{equation}
  A = \langle A \rangle - \frac{1}{2}\sum_{i=1}^3 \mu_{i,2} \partial_{x_i}^{2} \langle A\rangle +\mathcal{O}(\eta_i^4),
\label{eq:a}
\end{equation} 
where we omit the explicit dependence of $A$ for brevity.

Following a similar procedure, it is straightforward to derive an expression for the coarse-grained product of two continuous fields $A$ and $B$:
\begin{equation}
\langle A B\rangle = \langle A \rangle \langle B \rangle + \sum_{i=1}^3 \mu_{i,2} \partial_{x_i} \langle A\rangle \partial_{x_i} \langle B \rangle+\mathcal{O}(\eta_i^4) ,
\label{eq:ab}
\end{equation}
and the product of three continuous fields, $A$, $B$, and $C$:
\begin{equation}
\begin{aligned}
\langle A B C\rangle &= \langle A \rangle \langle B \rangle \langle C \rangle + \sum_{i=1}^3 \mu_{i,2} \left[ \partial_{x_i} \langle A\rangle \partial_{x_i} \langle B \rangle \langle C \rangle \right. \\
&\left.+ \partial_{x_i} \langle A\rangle  \langle B \rangle \partial_{x_i}\langle C \rangle +  \langle A\rangle \partial_{x_i} \langle B \rangle \partial_{x_i}\langle C \rangle\right] \\
&+\mathcal{O}(\eta_i^4) .
\end{aligned}
\label{eq:abc}
\end{equation}
Eq. (\ref{eq:ab}) represents a generalization of Eq (2.3) from Reference [\citen{margolin2009finite}] to a coarse-graining of the form shown in Eq. (\ref{eq:ave}). These relations simplify somewhat if we assume an isotropic coarse-graining, so that $\mu_{1,n}=\mu_{2,n}=\mu_{3,n}=\mu_n$. Defining the microscopic length scale $\ell\equiv\sqrt{\mu_{2}}$ and taking the macroscopic length scale $L$ to be much larger than $\ell$, we can define the small parameter $\eta\equiv\ell/L$. With these definitions we find that the above relations become:
\begin{equation}
\begin{aligned}
      A &= \langle A \rangle - \frac{\ell^2}{2} \nabla^2 \langle A\rangle +\mathcal{O}(\eta^4), \\
      \langle A B\rangle &= \langle A \rangle \langle B \rangle + \ell^2 \grad \LL A\RR\cdot \grad \LL B \RR+\mathcal{O}(\eta^4) , \\
      \langle A B C\rangle &= \LL A \RR \LL B \RR \LL C \RR + \ell^2 \left[ \grad \LL A\RR \cdot\grad \LL B \RR \LL C \RR \right. \\
&\left.+  \grad \LL A\RR  \LL B \RR \cdot\grad\LL C \RR +   \LL A\RR  \grad\LL B \RR \cdot\grad\LL C \RR\right] \\
&+\mathcal{O}(\eta^4) .
\end{aligned} 
\label{eq:FS}
\end{equation}
As we will demonstrate in the next section, Eqs. (\ref{eq:FS}) provide sufficient conditions to close the coarse-grained Madelung equations, representing an extension of the coarse-graining framework employed in previous work\cite{tanogami2021theoretical}. 

\section{Finite Scale Madelung Equations}
\label{sec:FS}

\subsection{Continuity Equation}
Using the averaging defined in Eq. (\ref{eq:ave}), we can write the continuity equation as:
\begin{align*}
\partial_t \langle \rho \rangle + \grad \cdot \langle \rho \textbf{u} \rangle &=0. 
\end{align*}
Defining the Favre average:
\begin{equation}
    \langle \langle\textbf{u} \rangle\rangle = \frac{\langle \rho \textbf{u} \rangle}{\langle \rho\rangle},
\end{equation}
we find that the coarse-grained continuity equation can be written as:
\begin{equation}
\partial_t \langle \rho \rangle + \grad \cdot \left( \langle \rho \rangle  \langle \langle\textbf{u} \rangle\rangle \right) =0.
\label{eq:mass}
\end{equation}
Therefore, we see that the mass continuity equation applies exactly to these averaged quantities, where the Favre averaged velocity plays the role of a macroscopic velocity field.
We can expand this Favre averaged velocity in terms of the finite scale averaged velocity using Eq. (\ref{eq:ab}) to obtain: 
\begin{equation}
\vvu = \LL \vu \RR + \frac{\ell^2}{\langle \rho \rangle} \grad\LL \rho\RR \cdot \grad \LL  \vu \RR +\mathcal{O}(\eta^4).
\end{equation}
These two averages are, in general, different. 

\subsection{Momentum Equation}
We now turn our attention to the momentum equation in Eqs. (\ref{eq:madelung}). Multiplying the microscopic version of the momentum equation by $\rho$, we find:
\begin{equation}
    \partial_t \left( \rho \vu\right) + \sum_{i=1}^3 \partial_{x_i}\left( u_i \rho \vu \right) = \frac{\hbar^2}{2m^2} \rho \grad \left( \frac{\nabla^2 \sqrt{\rho}}{\sqrt{\rho}}\right)- \frac{g}{m} \rho\grad\rho,
\label{eq:momentum1}
\end{equation}
where we have used the microscopic continuity equation to simplify the left hand side. We will now apply the finite scale coarse-graining procedure to Eq. (\ref{eq:momentum1}). Due to the complexity of some of these terms, more care is required in applying the procedure to this equation than in the case of Eq. (\ref{eq:mass}).  

The first term on the left hand side may be written exactly as:
\begin{align*}
\LL\partial_t \left( \rho \vu\right)\RR &= \partial_t \left( \LL\rho\RR \frac{\LL \rho\vu \RR}{\LL\rho\RR}\right)=\partial_t \left( \LL\rho\RR \vvu\right), \\
&=\vvu \partial_t \LL\rho\RR + \LL\rho\RR\partial_t\vvu ,\\
&=-\vvu\left[\vvu\cdot\grad\LL\rho\RR + \LL\rho\RR\grad\cdot\vvu \right] \\
&+ \LL\rho\RR\partial_t\vvu, \\
&=-\sum_{i=1}^3 \partial_{x_i} \left[ \LL\LL u_i\RR\RR \LL\rho\RR \vvu \right] \\
&+ \LL\rho\RR\left[ \partial_t\vvu + \vvu\cdot\grad \ \vvu\right],
\end{align*}
where we have used Eq. (\ref{eq:mass}) to eliminate the time derivative of $\LL\rho\RR$. Collecting terms and dividing by $\LL\rho\RR$ we find the exact momentum equation relating the coarse-grained quantities:
\begin{widetext}
\begin{equation}
\begin{aligned}
    \partial_t \vvu + \vvu\cdot\grad \ \vvu  &= \frac{\hbar^2}{2m^2} \LL\LL  \grad \left( \frac{\nabla^2 \sqrt{\rho}}{\sqrt{\rho}}\right) \RR\RR- \frac{g}{m} \LL\LL\grad\rho\RR\RR +\frac{1}{\langle\rho\rangle} \sum_{i=1}^3\left\{\partial_{x_i}  \left[ \LL\LL u_i\RR\RR \LL\rho\RR \vvu - \LL u_i\rho \vu \RR \right] \right\}.
\end{aligned}    
\label{eq:momentum-full}
\end{equation}
\end{widetext}
Notice that the left hand side is of the same form as in the original Madelung equations, but with $\vu$ replaced with $\vvu$. Notice, also, the appearance of three terms on the right hand side of the equation: a quantum stress term:
\begin{equation}
\textbf{S}_{Q} = \frac{\hbar^2}{2m^2}\LL\LL  \grad \left( \frac{\nabla^2 \sqrt{\rho}}{\sqrt{\rho}}\right) \RR\RR,   
\end{equation}
a stress term due to microscopic interactions:
\begin{equation}
\textbf{S}_{I} = \frac{g}{m}\LL\LL\grad\rho\RR\RR,   
\end{equation}
and a classical stress term:
\begin{equation}
\textbf{S}_{C} = \frac{1}{\langle\rho\rangle} \sum_{i=1}^3\left\{\partial_{x_i}  \left[ \LL\LL u_i\RR\RR \LL\rho\RR \vvu - \LL u_i\rho \vu \RR \right] \right\}. 
\label{eq:classical-stress}
\end{equation}
The quantum stress is simply the Favre average of the quantum pressure term in the Madelung equations, the interaction stress is, similarly, the Favre average of the term due to interactions. The classical stress term has the form of the divergence of a stress tensor. These terms are consistent with previous analyses of the Madelung equations\cite{tanogami2021theoretical}.
In principle, one could perform the averages in Eq. (\ref{eq:classical-stress}) for specific quantum systems to study how these terms depend on the macroscopic flow variables. However, such a study is beyond the scope of the current work. Instead, following the finite scale analysis for classical fluids, we will expand each term in Eq. (\ref{eq:classical-stress}) perturbatively in $\ell^2$, thus extending the unclosed hierarchy of moments previously studied to a closed set of equations for coarse-grained quantities.

Expanding the average triple product in the classical stress term, we find:
\begin{align*}
\LL u_i \rho \vu \RR &= \langle u_i \rangle \langle \rho \rangle \langle \vu \rangle +  \ell^2 \left[ \grad \langle u_i\rangle \cdot \grad \langle \rho \rangle \langle \vu \rangle \right.  \\
&\left.+ \grad \langle u_i\rangle  \langle \rho \rangle \cdot \grad \ \langle \vu \rangle +  \langle u_i\rangle \grad \langle \rho \rangle \cdot \grad \ \langle \vu \rangle\right] + \mathcal{O}(\eta^4) ,
\end{align*}
where we have simply used the identity in Eq. (\ref{eq:abc}). Similarly, the triple product of averages may be written as:
\begin{align*}
\LL\LL u_i \RR\RR  \LL\rho \RR \vvu &= \LL u_i \RR \LL \rho \RR  \LL\vu\RR  + \ell^2 \left[  \LL u_i\RR \grad\LL \rho\RR \cdot \grad \LL  \vu \RR \right. \\
&\left.+  \grad\LL \rho\RR \cdot \grad \LL  u_i \RR \LL\vu\RR \right] + \mathcal{O}(\eta^4) .
\end{align*}
Combining these results, we find that Eq. (\ref{eq:classical-stress}) may be written as:
\begin{align*}
\textbf{S}_{C}&= -\frac{1}{\langle\rho\rangle}\ell^2\sum_{i=1}^3\partial_{x_i}\left[ \LL\rho\RR \grad \LL\LL u_i\RR\RR \cdot\grad \ \vvu \right]  +\mathcal{O}(\eta^4),
\end{align*}
where we have replaced $\LL\vu\RR$ with $\vvu$ because they are equivalent at this order in $\ell^2$. Notice that, when expanded to leading order in $\eta=\ell/L$, the classical stress appears to be reminiscent of an artificial viscous stress term from computational fluid dynamics.  

To gain insight into the relative importance of each of the stress terms on the right hand side, we define a characteristic magnitude of the velocity field $U_0$ and a characteristic macroscopic density $\rho_0$. Combining the characteristic velocity with the macroscopic length scale $L$ we define the time scale $\tau=L/U_0$. With these scales defined, it is straightforward to arrive at the non-dimensionalized momentum equation:
\begin{equation}
\begin{aligned}
    \partial_{\tau} \tilde{\vu} + \tilde{\vu}\cdot\grad_{\boldsymbol{\xi}} \ \tilde{\vu}  &= \frac{\hbar^2}{2m^2 L^2 U_0^2} \tilde{\textbf{S}}_{Q}- \frac{g \rho_0}{m U_0^2} \tilde{\textbf{S}}_{I} - \frac{\ell^2}{ L^2 } \tilde{\textbf{S}}_{C},
\end{aligned}    
\label{eq:non-dim}
\end{equation}
where we define the normalized stress terms:
\begin{equation}
\tilde{\textbf{S}}_{Q} = L^3\LL\LL  \grad \left( \frac{\nabla^2 \sqrt{\rho}}{\sqrt{\rho}}\right) \RR\RR,   
\end{equation}
\begin{equation}
\tilde{\textbf{S}}_{I} = \frac{L}{\rho_0}\LL\LL\grad\rho\RR\RR,   
\end{equation}
and:
\begin{equation}
\tilde{\textbf{S}}_{C} = \frac{L^2}{\langle\rho\rangle U_0^2}\sum_{i=1}^3\partial_{x_i}\left[ \LL\rho\RR \grad \LL\LL u_i\RR\RR \cdot\grad \ \vvu \right].   
\end{equation}
We now see that the relative importance of each stress term is determined, in part, by the magnitudes of the three dimensionless parameters:
\begin{equation}
    \begin{aligned}
        q&=\frac{\hbar^2}{2m^2 L^2 U_0^2}, \\
        \gamma&=\frac{g \rho_0}{m U_0^2}, \\
        \eta&=\frac{\ell^2}{ L^2 }.
    \end{aligned}
    \label{eq:params}
\end{equation}
From these quantities we can see that classical and quantum stresses are comparable in magnitude when the coarse-graining scale is chosen to be:
\begin{equation}
    \ell\sim\ell_{Q} =\frac{\hbar}{\sqrt{2}m U_0}.
    \label{eq:scale}
\end{equation}
This sets a natural finite scale at which quasi-classical behavior should be expected. 

For a fluid with a macroscopic velocity of the order $U_0\sim 1$m/s, and $m\sim m_{p}=1.67262192\times10^{-27}$ kg, we find this coarse-graining scale to be $\ell_{Q}\sim 45$ nm. From this analysis, we conclude that when the interaction energy density is sufficiently low compared to the macroscopic kinetic energy, $U_0^2/\rho_0>>g/m$, and we are only concerned with behavior of the flow fields on scales such that $\ell>>\frac{\hbar}{\sqrt{2}m U_0}$, then the dynamics of this system may be described by a set of classical fluid equations. 

Another possible choice for $U_0$ is the characteristic particle velocity for an ideal gas with temperature $T$: $U_0=\sqrt{3k_BT/m}$. Inserting this into Eq. (\ref{eq:scale}) we find that the quasi-classical coarse-graining scale becomes: $\ell_{Q}=\sqrt{\frac{\hbar^2}{6mk_BT}}$, which is roughly a factor of 6 smaller than the thermal de Broglie wavelength: $\lambda=\sqrt{\frac{2\pi \hbar^2}{ mk_BT}}$. Therefore, choosing $\ell>>\lambda$ is a sufficient condition for classical behavior, when the interactions can be neglected, as one would expect from the correspondence principle.

\subsection{Vorticity Equation}
\label{sec:vorticity}
While the velocity appearing in the microscopic Madelung equations, Eq. (\ref{eq:madelung}), is by definition irrotational, $\curl\vu=0$, the same is not necessarily true for the Favre averaged velocity appearing in the finite scale Madelung equations, Eqs. (\ref{eq:mass}) and (\ref{eq:momentum-full}). In general, the vorticity of this coarse-grained field has the form:
\begin{align*}
    \w\equiv\curl\vvu&=\curl\frac{\LL \rho \vu \RR}{\LL\rho\RR}, \\
    &= \frac{\LL \grad\rho\times \vu\RR}{\LL\rho\RR} -\frac{\grad\LL\rho\RR \times \LL\rho\vu\RR}{\LL\rho\RR^2}, \\
    &=\frac{1}{\LL\rho\RR}\left[\LL \grad\rho\times \vu\RR-\grad\LL\rho\RR \times\vvu\right]. 
\end{align*}
This expression is exact and applies to the coarse-grained velocity field at all orders in the finite scale, $\ell$. There is no reason to expect that it should necessarily be zero, in general. Moreover, we can expand the right hand side to leading order in $\ell^2$ to arrive at:
\begin{equation}
\begin{aligned}
    \omega_{i}&=\frac{\ell^2}{\LL\rho\RR^2}\sum_{j,k,l}\epsilon_{ijk}\left[\LL\rho\RR\partial_{x_j}\partial_{x_l}\LL\rho\RR \right. \\
    &\left.- \partial_{x_j}\LL\rho\RR \ \partial_{x_l}\LL\rho\RR \right]\partial_{x_l}\LL u_k \RR + \mathcal{O}(\eta^4).
\end{aligned}
\label{eq:w-small-ell}
\end{equation}

In the next section we will consider a concrete example of a solution to the Madelung equations whose coarse-grained description possesses finite vorticity, thus demonstrating that vorticity is present in the finite scale description of even some irrotational flows. Before we construct that example, we will consider the equation describing the dynamics of the finite scale vorticity field. This can be obtained, in the same way as is done in classical hydrodynamics, by taking the curl of the momentum equation, Eq. (\ref{eq:momentum-full}):
\begin{equation}
\begin{aligned}
    \partial_t \w &+ \vvu\cdot\grad \ \w  = \w\cdot\grad \ \vvu - \w \ \grad\cdot\vvu \\
    &+\curl\textbf{S}_{Q}- \curl\textbf{S}_{I} +\curl\textbf{S}_{C}.
\end{aligned}    
\label{eq:vorticity-full}
\end{equation}
Note that, just as in classical hydrodynamics, the left hand side is the Lagrangian derivative of the vorticity field, that is the time derivative of the vorticity in the frame of the fluid. Also, just as in classical hydrodynamics, a vortex-stretching term, $\w\cdot\grad \ \vvu$, emerges on the right hand side, implying that vorticity will, generally, be enhanced when the velocity field increases along the direction parallel to $\w$. This phenomenon is thought to play an important role in the cascade of energy from larger to smaller length scales in classical turbulence. Since it arises naturally upon coarse-graining, it stands to reason that similar phenomena should be expected to play a role in macroscopic effective descriptions of quantum turbulence as well, even though vorticity is, strictly-speaking, absent in quantum fluids. The emergence of this phenomenon is consistent with Tanogami's previous work studying the energy cascade in coarse-grained quantum fluids\cite{tanogami2021theoretical} and lends further credence to his hypothesis.

Simplifications can be made if we expand this equation to leading order in the dimensionless parameters in Eq. (\ref{eq:params}), $q$, $\gamma$, and $\eta$. In this case, to leading order $\curl\textbf{S}_{Q}= \curl\textbf{S}_{I}=0$, so that:
\begin{equation}
\begin{aligned}
    \partial_t \w &+ \vvu\cdot\grad \ \w  = \w\cdot\grad \ \vvu - \w \ \grad\cdot\vvu \\
    &-\ell^2\curl\left\{\frac{1}{\langle\rho\rangle}\sum_{i=1}^3\partial_{x_i}\left[ \LL\rho\RR \grad \LL\LL u_i\RR\RR \cdot\grad \ \vvu \right]\right\} \\ &+\mathcal{O}(\eta^4).
\end{aligned}    
\label{eq:vorticity-LO}
\end{equation}
From this equation we see that at leading order in these parameters the evolution of the vorticity only depends on coarse-grained quantities and does not depend explicitly on any of the parameters appearing in the original microscopic description.

\section{Example Flow Field: A Line Vortex}
\label{sec:example}
To better understand the emergence of vorticity in the finite scale theory outlined in the previous section, we will now consider an example flow field which is chosen to be amenable to an exact analytic treatment while also exhibiting emergent vorticity when coarse-grained. 

With these motivations in mind, we choose the velocity field to be that associated with a classical line vortex and a density away from the vortex core given by a Gaussian distribution in three dimensions. At a distance $R$ from the vortex core, we give the density a simple scaling $\rho\sim R^{2n}$, for $n\in\mathbb{N}$. The wavefunction associated with these choices is given by:
\begin{equation}
    \psi_0=\frac{\left(\x_{\perp}\cdot\x_{\perp}\right)^{n/2}}{(2\pi)^{3/4}\sigma^{3/2}r_0^n}e^{-\frac{\x\cdot\x}{4\sigma^2}}e^{i\frac{m}{\hbar}\frac{\Gamma}{2\pi}\tan^{-1}\left(\frac{x_2}{x_1}\right)},
    \label{wavefunction-general-text}
\end{equation}
where $\textbf{x}_\perp$ lies in the plane perpendicular to the vortex line and $r_0$ is chosen to normalize the wavefunction:
\begin{equation}
    r_0=\sqrt{2}\sigma \left(n!\right)^{\frac{1}{2n}} .
\end{equation}
While this is not, in general, a solution of the Schr\"{o}dinger equation it is a valid initial wavefunction which would evolve in time according to the Schr\"{o}dinger equation and it is a relatively simple example of a wavefunction which captures some of the essential features of a quantized vortex, and whose coarse-grained hydrodynamic description can be shown to possess finite vorticity.

With this choice of wavefunction, the initial density and velocity fields are:
\begin{equation}
    \begin{aligned}
        \rho&=\frac{m\left(\x_{\perp}\cdot\x_{\perp}\right)^{n}}{(2\pi)^{3/2}2^n\sigma^{2n+3} n!}e^{-\frac{\x\cdot\x}{2\sigma^2}}, \\
        \vu&=\frac{\Gamma}{2\pi}\frac{-x_2\hat{x}_1+x_1\hat{x}_2}{\x_{\perp}\cdot\x_{\perp}},
    \end{aligned}
\end{equation}
where $\Gamma$ is the circulation around the line vortex. We note that, while this flow field does possess finite \textit{circulation}, given by $\Gamma$, it is actually irrotational in that $\curl\vu=0$ everywhere, except at the singular point $x_1=x_2=0$, as one can easily verify from the expression above\cite{batchelor2000introduction}.

To perform the coarse-graining procedure discussed in the previous section, we must choose a distribution to use as a filter. For convenience, we choose the Gaussian distribution:
\begin{equation}
    f(\x)=\frac{1}{(2\pi)^{3/2}\ell^3}e^{-\frac{\x\cdot\x}{2\ell^2}}.
\end{equation}
This form has a few advantages, the most important one for the current task is that the product $\rho f$ is fairly easy to integrate. In principle, other choices of $f$ may be more or less convenient for understanding other problems, but a complete characterization of this topic is beyond the scope of the current work. 

We can now evaluate the coarse-grained fields, $\LL\rho\RR$ and $\vvu$. The details of this calculation, for generic integer $n\geq 0$, are given in Appendix \ref{appendix}. For the coarse-grained density we find, Eq. (\ref{density}):
\begin{equation}
  \LL\rho\RR=  e^{-\frac{z^2}{2\ell^2}\left(1-\frac{\tilde{\sigma}^2}{\ell^2} \right)} \frac{e^{-\frac{R^2}{2\ell^2}}m \left(2\tilde{\sigma}^2 \right)^{n+3/2}}{(4\pi)^{3/2}\ell^32^n\sigma^{2n+3}}  M\left(n+1,1,\kappa^2 \right),
  \label{eq:density}
\end{equation}
and for the coarse-grained velocity we find, Eq. (\ref{velocity}):
\begin{equation}
    \LL\LL \vu\RR\RR=\frac{R \  \Gamma }{4\pi\ell^{2} }    \frac{M\left(n+1,2,\kappa^2\right)}{M\left(n+1,1,\kappa^2\right)} \ \hat{\theta},
    \label{eq:velocity}
\end{equation}
where $M(a,b,z)$ is Kummer's confluent hypergeometric function, $\tilde{\sigma}=\sigma\ell/\sqrt{\ell^2+\sigma^2}$, $\kappa^2=R^2\tilde{\sigma}^2/2\ell^4$, and $\hat{\theta}$ is the unit vector in the azimuthal direction around the vortex core.  

From Eq. (\ref{eq:velocity}) it is straightforward to arrive at the vorticity, Eq. (\ref{vorticity-general}):
\begin{widetext}
\begin{equation}
       \w = \hat{z} \frac{ \Gamma }{2\pi\ell^{2} }\left\{ 1 - \kappa^2\frac{M\left(n+1,2,\kappa^2\right)}{M\left(n+1,1,\kappa^2\right)} \left[1+ n\frac{M\left(n+1,2,\kappa^2\right)}{M\left(n+1,1,\kappa^2\right)}\right] \right\},
       \label{eq:vorticity}
\end{equation}
\end{widetext}
where $\hat{z}$ is in the direction along the vortex line. This demonstrates the emergence of vorticity under the coarse-graining procedure described in the preceding sections, and Eq. (\ref{eq:vorticity}) can be used to numerically compute the vorticity for arbitrary $n\in\mathbb{N}$. However, much insight can be gained from considering the special cases of $n=0$ and $1$.

As we show in Appendix \ref{n=0}, the coarse-grained variables for $n=0$ can be written as:
\begin{equation}
    \LL\rho\RR=  e^{-\frac{z^2+R^2}{2\ell^2}\left(1-\frac{\tilde{\sigma}^2}{\ell^2} \right)} \frac{m \left(2\tilde{\sigma}^2 \right)^{3/2}}{(4\pi)^{3/2}\ell^3\sigma^3}  ,
    \label{eq:density-0}
\end{equation}
\begin{equation}
    \LL\LL \vu\RR\RR=\frac{\Gamma }{2\pi }    \frac{1-e^{-\frac{R^2\tilde{\sigma}^2}{2\ell^4}}}{R\sigma^2}\left(\sigma^2 + \ell^2 \right) \ \hat{\theta},
    \label{eq:velocity-0}
\end{equation}
and
\begin{equation}
       \w = \hat{z} \frac{ \Gamma }{2\pi\ell^{2} }e^{-\frac{R^2\tilde{\sigma}^2}{2\ell^4}}.
       \label{eq:vorticity-0}
\end{equation}
From Eq. (\ref{eq:vorticity-0}) we see that the magnitude of the vorticity at the vortex core, $R=0$, is $\omega=\Gamma/2\pi\ell^2$. In the limit of $\ell\rightarrow0$ this diverges, consistent with the singular nature of the microscopic fields. Away from the vortex core, $R>0$, we see that $\lim_{\ell\rightarrow0}\omega=0$, consistent with the irrotational nature of the microscopic velocity field. However, we note that for $n=0$ the density does not vanish at the vortex core, and so this case is not as physically relevant as the $n=1$ case.

As we show in Appendix \ref{n=1}, the coarse-grained variables for $n=1$ can be written as:
\begin{equation}
    \LL\rho\RR= e^{-\frac{z^2+R^2}{2\ell^2}\left(1-\frac{\tilde{\sigma}^2}{\ell^2} \right)} \frac{m \left(2\tilde{\sigma}^2 \right)^{5/2}}{(4\pi)^{3/2}\ell^32\sigma^{5}}  \left(1+\frac{R^2\sigma^2}{2\ell^2\left(\ell^2+\sigma^2\right)}\right),
    \label{eq:density-1}
\end{equation}
\begin{equation}
    \LL\LL \vu\RR\RR=\frac{R \  \Gamma }{2\pi }    \frac{1}{2\ell^2+\tfrac{R^2\sigma^2}{\ell^2+\sigma^2}} \ \hat{\theta},
    \label{eq:velocity-1}
\end{equation}
and
\begin{equation}
       \w = \hat{z} \frac{ \Gamma }{\pi } \frac{ 2\ell^2}{\left(2\ell^2+\tfrac{R^2\sigma^2}{\ell^2+\sigma^2}\right)^2}.
       \label{eq:vorticity-1}
\end{equation}
At $R=0$, the vorticity is again $\w=\hat{z}\Gamma/2\pi\ell^{2}$, just as the $n=0$ case. However, for $R,\sigma>0$, and $\ell<<R,\sigma$, the vorticity becomes: 
\begin{equation}
       \w = \hat{z} \frac{ 2\Gamma }{\pi } \frac{\ell^2}{R^4}+\mathcal{O}(\ell^4).
       \label{eq:w-1:small-ell}
\end{equation}
Comparing this to the finite scale expansion in Eq. (\ref{eq:w-small-ell}), we confirm that the leading-order contributions for this case are proportional to $\ell^2$. This offers a concrete example of a vorticity field for which the finite-scale equations in Sec.~\ref{sec:FS} should capture the leading-order physics in small $\ell$.

\begin{figure}
\begin{center}
  \centering
\includegraphics[width=0.45\textwidth]{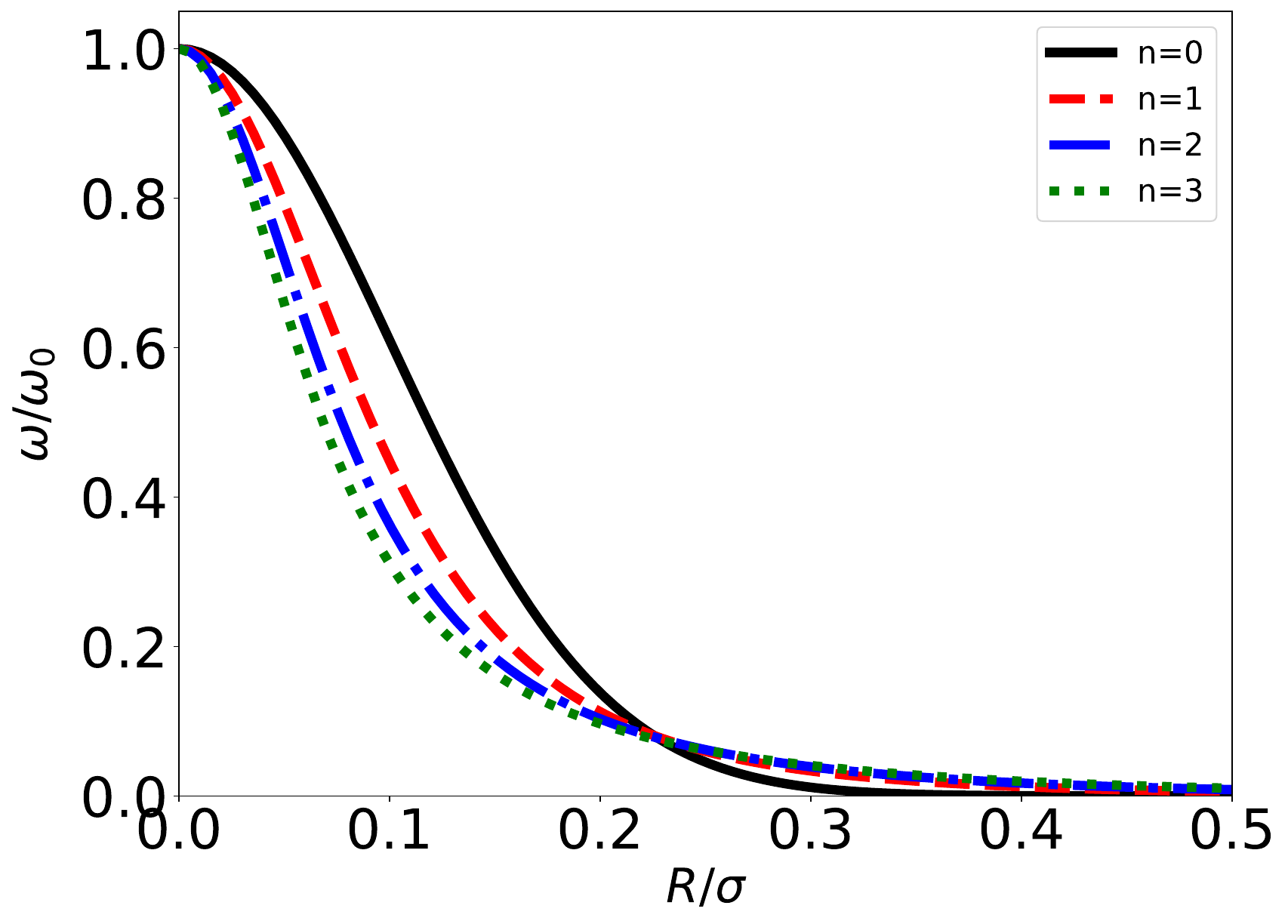}
\caption{Plot of emergent vorticity as given by Eq. (\ref{eq:vorticity}) as a function of radius from vortex center, $R$, normalized to the length scale $\sigma$. Vorticity is normalized to $\omega_0=\Gamma/2\pi\ell^2$, and $\ell/\sigma$ is chosen to be $1/10$ for illustrative purposes.}
\label{fig:w}
\end{center}
\end{figure}

Figure \ref{fig:w} shows a comparison of the emergent vorticity field, computed using the general formula, Eq. (\ref{eq:vorticity}), for $n=0,1,2,3$. Note that in all four cases, the vorticity is peaked at the vortex core and decreases to zero at large radii $R/\sigma$. Moreover, we see that the width of the peak becomes narrower as $n$ increases, and, correspondingly, the tail drops off more slowly. In this sense, we can think of the coarse graining procedure as smoothing out the singular behavior of the line vortex and distributing the circulation into a region with finite vorticity.

\section{Discussion and Conclusions}
\label{sec:conclusions}
In this paper we have derived a closed set of hydrodynamic equations for spatially coarse-grained quantum systems. The coarse-graining procedure we employed leads to an infinite hierarchy of filtered moments, equivalent to what has been discussed previously in the literature. This hierarchy is then closed using a generalization of finite scale theory which has been previously applied in the study of classical fluids. Consistent with previous results in classical fluids, as well as recent work on quantum fluids\cite{tanogami2021theoretical}, we find that the appropriate velocity field in the macroscopic fluid equations is a Favre averaged velocity field and that the coarse-graining gives rise to emergent viscous stresses in the momentum equations. After applying the finite-scale closure, these viscous stresses have the appearance of an artificial viscosity term, similar to those used in computational fluid dynamics. 
We have also shown that this macroscopic velocity field can possess an emergent vorticity, even though the microscopic velocity is irrotational. The equation governing the evolution of the vorticity was also derived and shown to possess the same vortex-stretching term that is present in classical fluids. Furthermore, expanding the stress terms to leading order in small parameters, we found that the equation governing the vorticity evolution depends only on coarse-grained quantities and has no dependence on the parameters describing the microscopic quantum model, consistent with its emergence at finite scales. 

The existence of finite vorticity was demonstrated explicitly in Sec.\ref{sec:example} using the specific example of an irrotational line vortex. It was shown that the emergent vorticity in that example is proportional to the vortex strength and that the coarse-graining acts to smooth out the singular behavior of the underlying microscopic velocity field and distribute finite vorticity through the fluid around the vortex. Importantly, when the density field is chosen to vanish at the vortex core, as it should in real systems, the leading order terms for the emergent vorticity  are quadratic in the coarse-graining length scale, exactly as they appear in the finite-scale theory presented in Sec.~\ref{sec:FS}. Therefore, we expect that the finite-scale closure used here should capture the vortex dynamics of real quantized vortices.

This analysis offers a different perspective on quantum fluids and the emergence of classical fluid dynamics from microscopic processes. It suggests that the main terms contributing to vorticity dynamics at large scales are, in fact, universal, in that they do not depend on the microscopic degrees of freedom. This has important implications because vorticity dynamics play a major role in the energy cascade from large to small length scales in classical turbulence. 

\begin{acknowledgements}
I would like to thank Alexander Balatsky, Roy Baty, and Len Margolin for insightful discussions and comments on various topics related to fluid dynamics, both quantum and classical. 
\end{acknowledgements}

\bibliography{finite_scale.bib}


\appendix

\section{Line Vortex Coarse-Graining}
\label{appendix}
In this appendix the closed form expressions for a coarse-grained line vortex are derived. Assume a wavefunction:
\begin{equation}
    \psi_0=\frac{\left(\x_{\perp}\cdot\x_{\perp}\right)^{n/2}}{(2\pi)^{3/4}\sigma^{3/2}r_0^n}e^{-\frac{\x\cdot\x}{4\sigma^2}}e^{i\frac{m}{\hbar}\frac{\Gamma}{2\pi}\tan^{-1}\left(\frac{x_2}{x_1}\right)},
    \label{wavefunction-general}
\end{equation}
where $r_0$ is chosen to normalize the wavefunction. For $n\in\mathbb{N}$:
\begin{equation}
    r_0=\sqrt{2}\sigma \left(n!\right)^{\frac{1}{2n}} .
\end{equation}

With this choice of wavefunction, the initial density and velocity fields are:
\begin{equation}
    \begin{aligned}
        \rho&=\frac{m\left(\x_{\perp}\cdot\x_{\perp}\right)^{n}}{(2\pi)^{3/2}2^n\sigma^{2n+3} n!}e^{-\frac{\x\cdot\x}{2\sigma^2}}, \\
        \vu&=\frac{\Gamma}{2\pi}\frac{-x_2\hat{x}_1+x_1\hat{x}_2}{\x_{\perp}\cdot\x_{\perp}},
    \end{aligned}
\end{equation}
where $\Gamma$ is the circulation around the line vortex.

For the following coarse-graining analysis, a Gaussian filter will be used, which we write as:
\begin{equation}
    f(\x)=\frac{1}{(2\pi)^{3/2}\ell^3}e^{-\frac{\x\cdot\x}{2\ell^2}}.
\end{equation}

We can now evaluate the coarse-grained fields, $\LL\rho\RR$ and $\vvu$. Starting with the density, we find:
\begin{align*}
    \LL\rho\RR&=\int d^3x' f\left(\x' \right)\rho\left(\x' + \x\right) , \\
    &= \frac{m}{(2\pi)^{3}\ell^32^n\sigma^{2n+3} n!} \int_{-\infty}^\infty dz'e^{-\frac{z'^2}{2\sigma^2}}e^{-\frac{z'^2+z^2-2zz'}{2\ell^2}} \\
    &\times\int d^2x'_\perp\left(\x'_\perp\cdot\x'_\perp\right)^{n} e^{-\frac{\x'_\perp\cdot\x'_\perp}{2\sigma^2}}e^{-\frac{\left(\x'_\perp-\x_\perp\right)\cdot\left(\x'_\perp-\x_\perp\right)}{2\ell^2}}, \\
    &= \frac{me^{-\frac{z^2}{2\ell^2}\left(1-\frac{\tilde{\sigma}^2}{\ell^2} \right)}\tilde{\sigma}\sqrt{2\pi}}{(2\pi)^{3}\ell^32^n\sigma^{2n+3} n!} e^{-\frac{R^2}{2\ell^2}} \\
    &\times\int_0^{2\pi} d\phi \int_{0}^{\infty} dr \ r^{2n+1} e^{-\frac{r^2}{2\tilde{\sigma}^2}}e^{\frac{rR\cos\phi}{\ell^2}}, \\
\end{align*}
where $\tilde{\sigma}=\frac{\sigma \ell}{\sqrt{\ell^2+\sigma^2}}$. Changing variables and using the definition of the modified Bessel functions $I_n$ we find:
\begin{equation}
\begin{aligned}
    \LL\rho\RR&= \frac{me^{-\frac{z^2}{2\ell^2}\left(1-\frac{\tilde{\sigma}^2}{\ell^2} \right)}\tilde{\sigma}\sqrt{2\pi}}{(2\pi)^{2}\ell^32^n\sigma^{2n+3} n!} e^{-\frac{R^2}{2\ell^2}} \left(2\tilde{\sigma}^2 \right)^{n+1} \\
    &\times\int_{0}^{\infty} dt \ t^{n} e^{-t}I_0\left( 2\kappa\sqrt{t}\right), \\
\end{aligned}
\label{eq:rho-bessel}
\end{equation}
where we define $\kappa\equiv R\tilde{\sigma}/\sqrt{2}\ell^2$. Note that the integral on the right hand side can be written exactly in terms of special functions\cite{gradshteyn2014table}:
\begin{equation}
\begin{aligned}
\int_0^\infty &x^{\mu-\frac{1}{2}}e^{-\alpha x}I_{2\nu}\left(2\beta\sqrt{x} \right)dx=\frac{\Gamma\left(\mu + \nu +\tfrac{1}{2} \right)}{\Gamma\left(2\nu +1 \right)} \\
&\times \frac{\beta^{2\nu}}{\alpha^{\mu+\nu+\tfrac{1}{2}}}  M\left(\mu+\nu+\tfrac{1}{2},2\nu+1,\frac{\beta^2}{\alpha}\right),
\end{aligned}
\label{integral}
\end{equation}
where $M$ is the confluent hypergeometric function, and $\text{Re}\left\{\mu+\nu+\tfrac{1}{2} \right\}>0$. 

It is straightforward to see that the integral in Eq. (\ref{eq:rho-bessel}) can be evaluated using Eq. (\ref{integral}) with the identification: $\mu=n+\tfrac{1}{2}$, $\alpha=1$, $\nu=0$, and $\beta=\kappa$. This yields:
\begin{equation}
  \LL\rho\RR=  e^{-\frac{R^2}{2\ell^2}}e^{-\frac{z^2}{2\ell^2}\left(1-\frac{\tilde{\sigma}^2}{\ell^2} \right)} \frac{m \left(2\tilde{\sigma}^2 \right)^{n+3/2}}{(4\pi)^{3/2}\ell^32^n\sigma^{2n+3}}  M\left(n+1,1,\kappa^2 \right).
  \label{density}
\end{equation}

Turning our attention to the Favre-averaged velocity field, we write the numerator as:
\begin{align*}
   \LL\rho\vu \RR &=\frac{m \Gamma e^{-\frac{z^2}{2\ell^2}\left(1-\frac{\tilde{\sigma}^2}{\ell^2} \right)}\tilde{\sigma}\sqrt{2\pi}}{(2\pi)^{4}\ell^3\sigma^3r_0^{2n}} \int d^2x_\perp' \left(\x'_\perp\cdot\x'_\perp\right)^{n-1} \\
    &\times e^{-\frac{\x_\perp'\cdot\x_\perp'}{2\sigma^2}}e^{-\frac{\left(\x_\perp'-\x_\perp\right)\cdot\left(\x_\perp'-\x_\perp\right)}{2\ell^2}} \left(-y'\hat{x} + x'\hat{y}\right), \\   
    &=\frac{m \Gamma e^{-\frac{z^2}{2\ell^2}\left(1-\frac{\tilde{\sigma}^2}{\ell^2} \right)}\tilde{\sigma}\sqrt{2\pi}}{(2\pi)^{4}\ell^3\sigma^3r_0^{2n}} \\
    &\times\int_0^{2\pi} d\phi \int_0^\infty dr \ r^{2n} e^{-\frac{r^2}{2\sigma^2}}e^{-\frac{r^2+R^2-2rR\cos\phi}{2\ell^2}} \\
    &\times\left[-\sin\left(\phi+\theta\right)\hat{x} + \cos\left(\phi+\theta\right)\hat{y}\right], \\ 
    &=\frac{m \Gamma e^{-\frac{z^2}{2\ell^2}\left(1-\frac{\tilde{\sigma}^2}{\ell^2} \right)}\tilde{\sigma}\sqrt{2\pi}}{(2\pi)^{4}\ell^3\sigma^3r_0^{2n}} e^{-\frac{R^2}{2\ell^2}} \\
    &\times \int_0^{2\pi} d\phi \int_0^\infty dr \ r^{2n} e^{-\frac{r^2}{2\tilde{\sigma}^2}+\frac{rR}{\ell^2}\cos\phi} \\
    &\times\left[-\sin\phi\hat{r} + \cos\phi\hat{\theta}\right], \\     
\end{align*}
where $\theta$ is defined by $\x_\perp\cdot\hat{x}=R\cos\theta$. Note that the term proportional to $\hat{r}$ vanishes after performing the $\phi$-integral. Dividing this result by $\LL\rho\RR$, and working through the algebra, we find:
\begin{align*}
  \LL\LL \vu\RR\RR&=\frac{   \Gamma }{2\pi \tilde{\sigma}  n! M\left(n+1,1,\frac{R^2\tilde{\sigma}^2}{2\ell^4} \right)}\frac{1}{\sqrt{2}} \\
  &\times\int_0^\infty dt \ t^{n-\tfrac{1}{2}} e^{-t} I_1\left( 2\kappa\sqrt{t}\right)  \ \hat{\theta}.
\end{align*}
The integral on the right hand side can be evaluated using the identity in Eq. (\ref{integral}), and identifying: $\mu=n$, $\alpha=1$, $\nu=\tfrac{1}{2}$, $\beta=\kappa$, leading to:
\begin{equation}
    \LL\LL \vu\RR\RR=\frac{R \  \Gamma }{4\pi\ell^{2} }    \frac{M\left(n+1,2,\kappa^2\right)}{M\left(n+1,1,\kappa^2\right)} \ \hat{\theta}.
    \label{velocity}
\end{equation}
Using this expression, we can write the vorticity as:
\begin{align*}
    \w &= \curl\LL\LL\vu\RR\RR= \frac{1}{R}\hat{z}\frac{\partial}{\partial R}\left(R \LL\LL u_\theta\RR\RR\right) , \\
&= \hat{z} \frac{ \Gamma }{4\pi\ell^{2} }\left[     \frac{M\left(n+1,2,\kappa^2\right)}{M\left(n+1,1,\kappa^2\right)} + \frac{\partial}{\partial R} R     \frac{M\left(n+1,2,\kappa^2\right)}{M\left(n+1,1,\kappa^2\right)}\right]. \end{align*}
This can be expressed exactly in terms of just $M\left(n+1,1,\kappa^2\right)$ and $M\left(n+1,2,\kappa^2\right)$ using the general relation from the NIST Digital Library of Mathematical Functions (13.3.15)\cite{DLMF}:
\begin{equation}
    \frac{dM(a,b,z)}{d z}=\frac{a}{b}M(a+1,b+1,z),
    \label{dM}
\end{equation}
together with the contiguous relations for $M(a,b,z)$\cite{DLMF}. While the analysis contained in this appendix only makes use of three of the contiguous relations, all six are included for completeness:
\begin{widetext}
\begin{equation}
    \begin{aligned}
(b-a) M(a-1,b,z) + (2a-b+z)M(a,b,z) -aM(a+1,b,z)&=0, \ \ (i) \\
b(b-1)M(a,b-1,z)+b(1-b-z)M(a,b,z)+z(b-a)M(a,b+1,z)&=0, \ \ (ii) \\
(a-b+1)M(a,b,z)-aM(a+1,b,z)+(b-1)M(a,b-1,z)&=0, \ \ (iii) \\
bM(a,b,z)-bM(a-1,b,z)-zM(a,b+1,z)&=0, \ \ (iv) \\
b(a+z)M(a,b,z)+z(a-b)M(a,b+1,z)-abM(a+1,b,z)&=0, \ \ (v) \\
(a-1+z)M(a,b,z) +(b-a)M(a-1,b,z) +(1-b)M(a,b-1,z)&=0. \,  \ \ (vi)
    \end{aligned}
    \label{M-relations}
\end{equation}
These six relations can be used to connect $M(a,b,z)$ to the four functions: $M(a\pm1,b,z)$ and $M(a,b\pm1,z)$. For the purposes of this appendix, we combine the $(vi)$ with $(v)$ to arrive at:
\begin{equation}
    M(a+1,b+1,z)=\frac{a-b}{a}M(a,b+1,z) + \frac{b}{a}M(a,b,z),
    \label{M:1}
\end{equation}
and also:
\begin{equation}
    M(a+1,b+1,z)=\frac{b}{z}\left[M(a+1,b,z) - M(a,b,z)\right].
    \label{M:2}
\end{equation}
Combining $(iii)$ with Eq. (\ref{M:2}) we find:
\begin{equation}
    M(a+1,b+1,z)=\frac{b(b-1)}{az}\left[M(a,b-1,z) - M(a,b,z)\right].
    \label{M:3}
\end{equation}
Using Eq. (\ref{dM}), together with Eqs. (\ref{M:1}) and (\ref{M:3}), it is straightforward to show that:
\begin{equation}
       \w = \hat{z} \frac{ \Gamma }{2\pi\ell^{2} }\left\{ 1 - \kappa^2\frac{M\left(n+1,2,\kappa^2\right)}{M\left(n+1,1,\kappa^2\right)} \left[1+ n\frac{M\left(n+1,2,\kappa^2\right)}{M\left(n+1,1,\kappa^2\right)}\right] \right\}.
       \label{vorticity-general}
\end{equation}
\end{widetext}
The general form for the coarse-grained vorticity, Eq. (\ref{vorticity-general}), can be used to evaluate the emergent vorticity for any vortex line described by the general form in Eq. (\ref{wavefunction-general}), where $\rho\sim R^{2n}$ near the vortex core. Notice that the general forms for the coarse-grained density, velocity, and vorticity, Eqs. (\ref{density}), (\ref{velocity}), and (\ref{vorticity-general}), are written in terms of just $M(n+1,1,\kappa^2)$ and $M(n+1,2,\kappa^2)$. 

Let us now consider two special cases, for which the confluent hypergeometric functions reduce to elementary functions. 

\subsection{Special Case: n=0}
\label{n=0}
The case of $n=0$ corresponds to the density field:
\begin{equation}
    \rho=\frac{m}{(2\pi)^{3/2}\sigma^3}e^{-\frac{\x\cdot\x}{2\sigma^2}}.
\end{equation}
This does not correspond to a physical quantized vortex, for which $\rho$ must vanish at the core; however, it is a simple and instructive example. In this case, the two confluent hypergeometric functions we need to evaluate are: $M(1,1,\kappa^2)$ and $M(1,2,\kappa^2)$. The first is simply given by:
\begin{equation}
    M(1,1,\kappa^2)=e^{\kappa^2},
    \label{eq:M11}
\end{equation}
which is easy to see from the definition of $M(a,b,z)$ in terms of the hypergeometric series:
\begin{align*}
    M(a,b,z)&=\sum_{k=0}^\infty\frac{(a)_k}{(b)_k}\frac{z^k}{k!},
\end{align*}
where $(a)_k$ and $(b)_k$ are Pochhammer symbols, where:
\begin{align*}
    (x)_k&=\frac{\Gamma(x+k)}{\Gamma(x)},
\end{align*}
so that:
\begin{align*}
    \frac{(a)_k}{(b)_k}&=\frac{\Gamma(a+k)}{\Gamma(a)}\frac{\Gamma(b)}{\Gamma(b+k)}.
\end{align*}
Note from this definition that $M(a,a,z)=e^z$. It is also straightforward to see:
\begin{align*}
    M(0,b,z)&=\sum_{k=0}^\infty\frac{\Gamma(k)}{\Gamma(0)}\frac{\Gamma(b)}{\Gamma(b+k)}\frac{z^k}{k!}, \\
    &=1 + \frac{1}{\Gamma(0)}\sum_{k=1}^\infty\frac{\Gamma(k)\Gamma(b)}{\Gamma(b+k)}\frac{z^k}{k!}, \\
    &=1.
\end{align*}
With Eq. (\ref{eq:M11}) and relation $(vi)$ from Eqs. (\ref{M-relations}), it is straightforward to find:
\begin{equation}
        M(1,2,\kappa^2) =\frac{e^{\kappa^2}-1}{\kappa^2}.  
        \label{eq:M12}
\end{equation}

With Eqs. (\ref{eq:M11}) and (\ref{eq:M12}) we see that the coarse-grained variables for $n=0$ are:
\begin{equation}
    \LL\rho\RR=  e^{-\frac{z^2+R^2}{2\ell^2}\left(1-\frac{\tilde{\sigma}^2}{\ell^2} \right)} \frac{m \left(2\tilde{\sigma}^2 \right)^{3/2}}{(4\pi)^{3/2}\ell^3\sigma^3}  ,
    \label{density-0}
\end{equation}
\begin{equation}
    \LL\LL \vu\RR\RR=\frac{\Gamma }{2\pi }    \frac{1-e^{-\frac{R^2\tilde{\sigma}^2}{2\ell^4}}}{R\sigma^2}\left(\sigma^2 + \ell^2 \right) \ \hat{\theta},
    \label{velocity-0}
\end{equation}
and
\begin{equation}
       \w = \hat{z} \frac{ \Gamma }{2\pi\ell^{2} }e^{-\frac{R^2\tilde{\sigma}^2}{2\ell^4}}.
       \label{vorticity-0}
\end{equation}
Notice that, for $R=0$ the vorticity is $\w=\hat{z}\Gamma/2\pi\ell^{2}$ and, hence, diverges as $\ell\rightarrow0$. This is expected due to the singular nature of vorticity in the microscopic problem. \\

\subsection{Special Case: n=1}
\label{n=1}
The case of $n=1$ corresponds to the density field:
\begin{equation}
    \rho=\frac{m\left(\x_{\perp}\cdot\x_{\perp}\right)}{(2\pi)^{3/2}2\sigma^{5} }e^{-\frac{\x\cdot\x}{2\sigma^2}}.
\end{equation}
This corresponds to a more physical representation of quantized vortex, for which $\rho\sim R^2$ near the vortex core. In this case, the two confluent hypergeometric functions we need to evaluate are: $M(2,1,\kappa^2)$ and $M(2,2,\kappa^2)$. The second of these is clearly:
\begin{equation}
    M(2,2,\kappa^2)=e^{\kappa^2}.
    \label{eq:M22}
\end{equation}
To find $M(2,1,\kappa^2)$ we can again use relation $(vi)$ from Eqs. (\ref{M-relations}) to find:
\begin{equation}
    M(2,1,\kappa^2)=(1+\kappa^2)e^{\kappa^2}.
    \label{eq:M21}
\end{equation}

With Eqs. (\ref{eq:M22}) and (\ref{eq:M21}) we see that the coarse-grained variables for $n=1$ are:
\begin{equation}
    \LL\rho\RR= e^{-\frac{z^2+R^2}{2\ell^2}\left(1-\frac{\tilde{\sigma}^2}{\ell^2} \right)} \frac{m \left(2\tilde{\sigma}^2 \right)^{5/2}}{(4\pi)^{3/2}\ell^32\sigma^{5}}  \left(1+\frac{R^2\sigma^2}{2\ell^2\left(\ell^2+\sigma^2\right)}\right),
    \label{density-1}
\end{equation}
\begin{equation}
    \LL\LL \vu\RR\RR=\frac{R \  \Gamma }{2\pi }    \frac{1}{2\ell^2+\tfrac{R^2\sigma^2}{\ell^2+\sigma^2}} \ \hat{\theta},
    \label{velocity-1}
\end{equation}
and
\begin{equation}
       \w = \hat{z} \frac{ \Gamma }{\pi } \frac{ 2\ell^2}{\left(2\ell^2+\tfrac{R^2\sigma^2}{\ell^2+\sigma^2}\right)^2}.
       \label{vorticity-1}
\end{equation}
At $R=0$, the vorticity is $\w=\hat{z}\Gamma/2\pi\ell^{2}$, just as the $n=0$ case. However, for $R,\sigma>0$, and $\ell<<R,\sigma$, the vorticity becomes: 
\begin{equation}
       \w = \hat{z} \frac{ 2\Gamma }{\pi } \frac{\ell^2}{R^4}+\mathcal{O}(\ell^4).
       \label{vorticity-1:small-ell}
\end{equation}
Comparing this to the finite scale expansion presented in the main text, we confirm that the leading-order contributions to the vorticity go as $\sim\ell^2$.

\end{document}